\documentclass[10pt]{llncs}

\usepackage{amssymb}

\usepackage{url} 

\title{On the specification and verification of atomic swap smart contracts\thanks{Version of Sep 10, 2018.}
} 

\author{Ron van der Meyden\\
meyden@cse.unsw.edu.au}

\institute{UNSW Sydney}

\newcommand{\semc}{\hspace{1pt}\mbox{:}\hspace{1pt}} 
\newcommand{\holds}{\mathtt{holds}}
\newcommand{\rimp}{\Rightarrow}
\newcommand{\nec}{\Box} 
\newcommand{\plays}{~\mathtt{plays}~}
\newcommand{\awaiting}{~\mathtt{Awaiting}~}
\newcommand{\Eventually}{\mathtt{Eventually}~}
\newcommand{\Until}{~\mathtt{Until}~}
\newcommand{\Always}{\mathtt{Always}~}
\newcommand{\atlop}[1]{\langle\hspace{-2pt}\langle #1 \rangle \hspace{-2pt}\rangle}

\newcommand{\recoverable}{\mathit{Recoverable}}
\newcommand{\Swap}{\mathit{Swap}}

\begin{document} 

\maketitle

\begin{abstract} 
Blockchain systems and smart contracts provide ways to securely implement multi-party transactions without the 
use of trusted intermediaries, which currently underpin many commercial transactions. However, 
they do so by transferring trust to computer systems, raising the question of whether code can be trusted. 
Experience with high value losses resulting from incorrect code has already shown that formal verification 
of smart contracts is likely to be beneficial. This note investigates the specification and verification of a simple 
form of multi-party transaction, \emph{atomic swaps}. 
It is argued that logics with the ability to express properties of strategies of players in a multi-agent setting 
are conceptually useful for this purpose, although ultimately, for our specific examples, the less expressive setting of temporal logic 
suffices for verification of concrete implementations. This is illustrated through a number of examples 
of the use of a model checker to verify atomic swap smart contracts in on-chain and cross-chain settings. 
\end{abstract} 

\section{Introduction} 

Many commercial transactions presently make use of intermediaries, independent of the interacting parties. 
The role of these intermediaries is to overcome deficits of trust between the transacting parties, ensuring that 
none are able to act in a way that causes the others losses due to deviations from agreed terms of exchange. 

One of the main potential benefits of blockchain technology is its  ability to eliminate the use of trusted intermediaries in 
multiparty interactions, enabling parties to engage in reliable peer-to-peer transactions without having to place
trust in third parties. The ability to ensure atomicity of transactions composed of multiple independent exchanges is one 
significant example of this capability. 

While blockchain and smart contract solutions have the attractive property of being able to enforce such atomic transactions, 
they do come with risks of their own. Trust placed in a person is not in fact eliminated, rather, it is 
replaced by trust in the correct behaviour of the smart contract code. In effect, the fact that there is an intermediary in the transaction is  not
changed, but a human intermediary is replaced by an intermediary composed of smart contract code executing on a blockchain platform. 

Trust in code is not always warranted: code can be 
complex, and behave in unexpected ways. This is  particularly the case when, as in the case of smart contracts on the blockchain, code is open to 
inspection,  and malicious agents may interact with it. 
There have already been a number of instances where large losses have been incurred as a result of poorly written smart contracts, 
that were vulnerable to attacks enabling the attackers to either steal assets \cite{DAO-NYT}, or make them unavailable \cite{ParityPM} 
to the rightful owners. There is therefore a need for smart contracts, which may control large amounts of monetary value, 
to be subjected to rigorous quality assurance before trust is  warranted. 

The strongest form of quality assurance used in computer science uses \emph{formal methods}, involving the use of 
precise, mathematically grounded models  to rigorously represent specifications and 
implementations, and to \emph{prove} that the implementations meet the specifications. 
To ensure that the reasoning in these proofs is correct, they are represented in formal logics, and 
reasoning steps are checked (or even automatically constructed) using software tools. While such tools 
may themselves contain errors, the level of attention to detail that they provide is far beyond human capacity for 
tedious checking of large volumes of reasoning steps, and the methodology has been highly successful in 
identifying errors that had escaped human attention. 

There is significant interest in the blockchain community in the application of formal methods to blockchain systems and smart contracts. 
The first step in such an application is to develop mathematically precise specifications of the desired behaviour and properties of 
the systems to be developed. In this note, we conduct a case study of one of the simplest cases of a smart contract, in order to develop an understanding of 
the types of specifications that are required in this area. 

Specifically, we consider the case of an atomic swap of digital assets. The scenario is very simple: Alice and Bob each own an asset, and they would like
to exchange ownership. 
We explain this scenario informally, and sketch code for a simple smart contract implementing an atomic swap in Section~\ref{sec:swap}. 
We then consider a number of potential approaches for formal specifications against which such code might be judged for correctness. In Section~\ref{sec:prepost} we consider the 
suitability of pre-condition/post-condition specifications. Section~\ref{sec:atl} argues that a richer form of specification captures the intended behaviour 
better, by making use of logics that contain constructs relating to \emph{strategies} played by agents in a multi-agent setting. 
We argue that these constructs provide a conceptually useful high level language in which to express issues of concern in smart contract specification, 
but also that they have some limitations, and may ultimately be eliminable when it comes to verifying concrete implementations. 

We illustrate this in Section~\ref{sec:mc}, where we show that it is possible to automatically verify that a concrete implementation 
of atomic swap satisfies its specification using just temporal logic model checking, without making use of strategic operators. 
In Section~\ref{sec:ccswap}, we go on to apply the methodology of Section~\ref{sec:mc} to a harder problem: cross-chain atomic swaps, in which the 
assets to be swapped reside on different blockchains.  Section~\ref{sec:concl} concludes with a discussion of 
questions for future consideration and related work. 

\section{Atomic Swaps} \label{sec:swap}

Suppose Alice holds an asset $a$ and Bob holds an 
asset $b$. They would like to execute a swap, in order to reach a state where Alice holds $b$ and Bob holds $a$. 
The problem that this presents is that one of the parties may not behave as expected. If Alice first transfers $a$ to Bob, then 
she is at risk that Bob will not uphold his end of the bargain, and transfer $b$ to Alice. Alice would then face the loss of $a$ without
compensation. There is a similar problem if Bob first makes the transfer to Alice. (Alice and Bob's recourse to the courts to enforce the exchange contract may not be effective in a blockchain setting, where Alice and Bob may know each others' identity only via their public keys, in which case they would not even know who should be served with a legal claim.) 
 
One commercially applied solution to this problem is for Alice and Bob to make use of an intermediary known as an ``escrow agent". 
We will call this agent ``Esther" here.  Alice and Bob each transfers their asset to Esther, and once she has received 
both $a$ and $b$, Esther transfers $b$ to Alice and $a$ to Bob. This solution no longer requires that Alice and Bob trust each other, 
but it does require that both trust that Esther will act in their best interests. Were she malicious, Esther could cause many forms of harm. 
She could abscond with both Alice and Bob's assets herself, transfer $a$ to Bob before receiving $b$ from Bob (enabling Bob to abscond with both assets), or transfer just one of the assets and retain the other herself. Other more subtle forms of malice might be possible on Esther's part, like retaining the assets for an extended period before transferring them in order to collect part of their revenue streams for herself. 
This solution is only effective if Alice and Bob can justifiably believe that Esther is not malicious in any of these ways. Finding a person who can be trusted in this way may be a challenge, particularly if Alice and Bob live far from each other. Even if they can find an appropriate Esther, she is 
only human, so might still behave maliciously even if she has been trustworthy in the past. 

The promise of blockchain and smart contract technology for this problem (at least for digital, or digitally represented, assets)
is that they enable Esther to be replaced by a computer program
that is in-corruptible and does not act with malicious intent. Both Alice and Bob can carefully inspect this program and form the justified belief that it behaves in exactly the way that they were expecting of Esther. If this program runs on blockchain platform that Alice and Bob trust will reliably execute this programs exactly according to its specified semantics, then they can  be satisfied that their transaction will be executed without risk of loss. 

Figure~\ref{fig:sc1} shows the code for such an ``escrow" smart contract. Rather than use any specific smart contract programming language, we write psuedo-code.
Intuitively, the code consists defines a collection of functions that may be called by any agent. The function \verb+initialise+ is called automatically when the 
smart contract is initiated on the blockchain, and sets up initial values of the local  variables \verb+depositedA+, \verb+depositedB+, which are intended to
represent whether Alice and Bob, respectively, have transferred the expected asset to the smart contract. We assume that \verb+initialise+ can only be called at initiation, and has null effect  if anyone attempts to call it at any other time.

 \begin{figure} 
 \begin{verbatim} 
Contract Escrow 
{  
depositedA, depositedB : Bool 

initialise { depositedA := False ;  depositedB := False }  
 
depositA  { if sender = Alice and value = a then depositedA := True }  
 
depositB  { if sender = Bob and value = b then depositedB := True }  

finalise { if depositedA and depositedB 
               then { depositedA := False; depositedB := False;  
                      send(a,B); send(b,A) } }
 
cancelA { if depositedA then { depositedA := False; send(a,A) }  }  

cancelB { if depositedB then { depositedB := False; send(b,B) }  }  
}
 \end{verbatim} 
 \caption{\label{fig:sc1}An atomic exchange smart contract} 
 \end{figure}

The function \verb+depositA+ allows Alice to transfer the asset $a$ to the smart contract. 
The keyword \verb+sender+ denotes the agent who is calling a function. We assume that a function call may be made with an asset attached - the effect of 
attaching an asset with a function call is for the asset to be transferred to the smart contract. The smart contract code may check the value of the attached 
asset using the keyword \verb+value+. Thus, the code for \verb+depositA+ checks that the function has been called by Alice and has asset $a$ attached. If
so, then Alice is credited with the deposit by setting \verb+depositedA+ to be true, otherwise this function call has no effect. There is a similar function \verb+depositB+ allowing Bob to deposit his asset. 

If all goes well, and both assets have been deposited, then the function \verb+finalise+ may be called by any agent (but presumably, will be called by either Alice or Bob, since noone else stands to gain anything by doing so) and has the effect of transferring the two assets back to Alice and Bob in such a way as to effect the swap. 

It may happen however, that Alice deposits her asset, but Bob does not. This would be a problem for Alice, since the asset is then no longer under her control, but under the control of the smart contract, but she has received nothing in return. To protect Alice against the eventuality that Bob does not deposit his asset (or does not do some in a timely fashion), there is a function \verb+cancelA+ that enables Alice to cancel her participation in the exchange and recover her asset $a$. A similar function is available to Bob. 

We abstract cryptographic concerns from our presentation - typically Alice and Bob would be represented via their public keys, and their instructions to perform actions in the smart contract would be sent as cryptographically signed messages.  Note that  blockchain systems do not provide guarantees with respect to the order of execution of messages submitted for processing by the agents - this is under the control of network operators (who are typically called ``miners'' in the area.) Smart contract code is typically \emph{reactive}, i.e., the state of the smart contract changes only when some agent (or another smart contract) sends it a message - there are no spontaneous events such as events triggered by timers.

The escrow smart contract is relatively short, but it already displays some complexity, with several functions maintaining an internal state that consists of assets held by the contract, as well as variables used for book-keeping of those assets. One might wonder if there are any unexpected interactions between these functions, and whether, in combination, they capture the possible behaviours that Alice and Bob expect of their interaction. The code already has some implicit invariants that underly its correctness. For example, 
why it is safe for the send operations to be performed  in $\mathtt{finalise}$, $\mathtt{sendA}$ and $\mathtt{sendB}$? The reason is that the functions interact in such a way as to ensure that the following statement is an invariant that holds in all states reachable from any sequence of calls: if $\mathtt{depositedA}$, then the smart contract holds asset $a$, 
and similarly, if $\mathtt{depositedB}$, then the smart contract holds asset $b$. Even if one considers this obvious for this contract, for even a marginally more complex contract, it would be desirable to have an independent \emph{specification} against which the code could be evaluated for correctness. 

\section{Pre-condition/Post-condition specifications} \label{sec:prepost} 

One commonly used form of formal specification of code is the use of \emph{pre-conditions} and \emph{post-conditions}. In the \emph{refinement calculus} \cite{Morgan94,BackWright} 
notation, these can be written using statements in the form $[pre,post]$, where $pre$ and $post$ are formulas that assert some property of the state of the system. The intuitive meaning of this statement is that it is a specification of a program. A concrete 
program $P$ \emph{implements} (or \emph{refines}) this statement if, whenever it is executed from an initial state satisfying the pre-condition $pre$,
at termination the program is in a state satisfying the post-condition $post$. 
There are weak and strong version of this semantics: in the weak (partial completeness) semantics, termination is not guaranteed, 
in the strong, the program is required to terminate whenever started at a state satisfying $\phi$. 
 
 It would appear at first that this specification approach can be used to specify the Escrow smart contract. Write $\holds(x,y)$ to mean that agent $x$ holds asset $y$. 
 Then (abbreviating Alice and Bob respectively as $A$ and $B$) we could try to specify the goal of Alice and Bob's exchange by the statement 
 \begin{equation} [\holds(A,a) \land \holds(B,b), ~\holds(A,b) \land \holds(B,a)] 
\label{prepost} 
 \end{equation}   
which is satisfied by a  program that  (provided it is started in the correct initial state) guarantees on termination that the assets have been swapped. 

Unfortunately, this specification is too strong to be implemented in a blockchain setting. Pre-condition/Post-condition specifications are applicable to settings with a 
single ``thread'', or locus of control, and this is not the case on the blockchain, where we need to consider not only the execution of smart contract  code, but also the actions of other agents in the system, in this case Alice and Bob. Evidently, action by both Alice and Bob is required to  bring the smart contract of Figure~\ref{fig:sc1} to the desired final state, 
and there is nothing that the contract can  do on its own to ensure that Alice and Bob each execute their part of the deal. Indeed, in order to apply precondition/post-condition specifications, we first need to clarify what we mean by  termination in such a multi-agent setting.  
For the moment, we take a terminated state to be one reached by executing an action called ``finalise'', and assume that the other actions are disabled 
after this function has been called. (This can easily be programmed into the smart contract by adding a 
variable $\mathtt{done}$ that is initially false and set true by $\mathtt{finalise}$ when this is called in a state where $\mathtt{depositedA}$ and $\mathtt{depositedB}$ hold.)

We could attempt to interpret the specification 
(\ref{prepost}) with the weak (partial completeness) semantics in order to cover the possibility that Alice or Bob does not play their part, and treat such a behaviour as a 
case of non-termination. However, this is also unsatisfactory. Consider a variant of the contract of Figure~\ref{fig:sc1}, in which we delete the 
actions $\mathtt{cancelA}$ and  $\mathtt{cancelB}$. This satisfies the weak interpretation of specification~\ref{prepost}, since now the 
only way to reach the final state is for both Alice and Bob to deposit their assets and then call $\mathtt{finalise}$. Failure by either Alice or Bob to play their part is treated as
non-termination, and the specification has nothing to say about this eventuality. However, from Alice and Bob's perspective, this is a fatal deficiency of 
the specification: this variant of their contract allows a run in which Alice deposits her asset, but Bob does not, with the effect that Alice has forever 
lost control of her asset, without compensation. 

Since it seems to be necessary to accommodate the possibility that the implementation terminates without the exchange having occurred, one 
might be tempted to weaken the specification to the following
 $$ \left[\holds(A,a) \land \holds(B,b), 
 ~\begin{array}{l} (\holds(A,b) \land \holds(B,a)) \\ 
                             \lor (\holds(A,a) \land \holds(B,b))\end{array} \right] ~,$$  
which allows that on termination, the system has reverted to its original state. However, this also does not capture our intent for the system. 
There is a very simple program ($skip$)  which implements this specification just by doing nothing!  Clearly, this is not what Alice and Bob intend: 
they do wish to engage in the exchange, but to do in a way that protects them from malicious behaviour by the other. At the very least, some additional 
constraint would need to be included in the formal specification to capture this. It is not at all clear how to express Alice and Bob's preference within 
a pre-condition/post-condition specification.

\section{Logics with Quantification over Strategies } \label{sec:atl} 

\emph{Temporal logics} \cite{Emerson} contain operators that express properties of a world whose state changes over time. They have been found useful for specifying 
and reasoning about concurrent systems, and are widely used for verification of such systems, particularly in the context of computer hardware development. 
There are several variants of such logics: \emph{linear time} logics are concerned with what happens in a single history, and \emph{branching time} logics 
take the view that time has a tree-like structure, with multiple \emph{possible} futures at each instant. 

Specification of concurrent settings involving multiple, possibly adversarial agents has been addressed in some types of logic called 
\emph{alternating temporal logic} \cite{ATLJACM}, and the more general \emph{strategy logic} \cite{CHP10}. These are extensions of branching time temporal logics. 
The semantic setting for these logics is an environment, called a \emph{concurrent game structure}  in the literature, 
in which some set of agents are each equipped with a  set of actions that they may perform. At each moment of time, agents that are scheduled to act
select one of their actions to perform, and a concurrent game structure describes how the selected actions cause a
state transition. 

A \emph{strategy} for an agent is a rule for selecting an action of the agent in each possible state of the system. 
For example, the set of conditions in Figure~\ref{fig:stratA} give a strategy for Alice in the environment created by the atomic exchange contract of 
Figure~\ref{fig:sc1}. Here the construct $\mathtt{do} \ldots \mathtt{od}$ is a nondeterministic loop statement, consisting of a set of clauses $C \rightarrow A$. 
The statement executes by repeatedly performing one of the actions $A$ for which the condition $C$ is true. If there are several possible such actions, the choice is 
made non-deterministically. 

\begin{figure} 
 \begin{verbatim}  
do
  holds(A,a) ->  depositA; 
  depositedA and depositedB -> finalise;  
  otherwise -> skip 
od
 \end{verbatim} 
 \caption{\label{fig:stratA}A strategy for Alice} 
 \end{figure} 

Alternating temporal logic introduced a specification construct $\atlop{G}\phi$, in which $G$ is a set of agent names, and 
$\phi$ is formula. Intuitively, the formula $\atlop{G}\phi$ says that the group of agents $G$ has a strategy that guarantees, whatever the 
other agents in the system do, that the formula $\phi$ will hold. The formula $\phi$ states some property of the future. 
For example, $\phi$ could be the formula $\Eventually \psi$, which holds if $\psi$ is true at some time in the future. 

Using this construct, we can capture more of the requirements discussed above. 
Write $I$ for the initial condition $\holds(A,a) \land \holds(B,b)$ and $F$ for the desired final condition $\holds(A,b) \land \holds(B,a)$. 
Then the formula 
\[ \atlop{A,B} ( I \rimp  \Eventually F ) \]
says ``Alice and Bob have a strategy that, if started in a state satisfying $I$,  guarantees that (whatever any other agent does) 
eventually  $F$." 

We can also express the idea that Alice should be able to recover her asset by some means (even if Bob is not cooperating) 
using the formula 
\[\atlop{A} \Eventually \holds(A,a) ~.\] 
This says ``Alice has a strategy that ensures (whatever any other agent 
does) that eventually Alice holds asset $A$''.  Obviously, this would be true trivially (without Alice having to do anything) at any time that Alice already holds
$a$, but it also ensures that Alice has some course of action that guarantees that she recovers $a$ in case, e.g., $a$ has been placed under the control of 
a smart contract. 

However, this latter requirement may not be realisable. In a blockchain implementation of the smart contract of Figure~\ref{fig:sc1}, Alice's strategy for 
recovering her asset (after she has already called $\mathtt{depositA}$), would be to send  message 
invoking action $\mathtt{cancelA}$. This message will take some time to spread through the miner network, and it 
may take some additional time before a miner chooses to incorporate it into a block. In the mean time, Bob may have been able to call 
$\mathtt{finalse}$ (preceded by  $\mathtt{depositB}$), and effect the swap. Miners may have chosen to prioritise Bob's messages. This means that the action actually under 
Alice's control (sending the request) is not guaranteed to  eventually make $\holds(A,a)$ true. However, in the event that 
Bob ``wins'' this race, the contract does result in $\holds(A,b)$ being true. From Alice's point of view,  this was her initial objective 
in doing business with Bob, so she should be satisfied, even if her attempt to cancel failed. 
(The point of the cancellation facility is to prevent her asset being locked in the contract  as a result of  Bob's potential malice,)
This suggests that the weaker formula 
\[\atlop{A} \Eventually (\holds(A,a) \lor  \holds(A,b)) \] 
is a more  satisfactory way to express Alice's ability to recover from malicious 
activity by Bob: it may be paraphrased as ``Alice may recover \emph{an} asset". A symmetric formula 
\[\atlop{B} \Eventually (\holds(B,b) \lor  \holds(B,a)) \] 
expresses Bob's ability to recover an asset. We write $\recoverable_A$ and $\recoverable_B$ for these formulas, respectively. 

Let us now combine these pieces, with some qualification. We do not require Alice and Bob to have the ability to recover their assets in all possible situations. 
For example, if Alice decides not to cooperate with Bob, and transfers her asset to Carol instead, then she then has no right to expect to recover 
her asset. Alice needs to be protected from Bob's malice only so long as she is engaged in her interaction with Bob. This suggests combining 
our requirements into a single formula as follows: 
\[
\atlop{A,B}
\left ( I \rimp 
   \left (\begin{array}[c]{l}
              \Eventually F~ \land \\
            \left ( 
                       \recoverable_A~ \land 
                       \recoverable_B
                     \right)
             \awaiting F  
\end{array} 
  \right)
\right)
\]
We use here a binary temporal construction $\alpha \awaiting \beta$, the meaning of which is that formula $\alpha$ continues to hold 
up to the moment that formula $\beta$ holds. It allows that $\beta$ never holds, in which case $\alpha$ holds forever.\footnote{Expressing this in 
linear time temporal logic, $\alpha~ \mbox{Awaiting} ~\beta$ is $(\alpha \Until \beta) \lor \Always (\alpha \land \neg \beta)$, which says that either 
$\alpha \Until  \beta$ (meaning $\beta$ holds at some time in the future, and $\alpha$ holds at every moment up until that time), or \
$\Always (\alpha \land \neg \beta)$ (meaning that $\alpha$ is true and $\beta$ false, now and at all times in the future).
We do not use the stronger formula $\alpha \Until \beta$ here to allow for easier comparison with a later formula, 
which has contexts where we cannot guarantee that  $\beta$ will ever hold.}  

However, this formula is still not quite what we want. It says that Alice and Bob retain the ability to recover an asset, so long 
as both are playing the strategy intended to eventually result in the final condition $F$, up to the moment that this final condition holds. 
But what of a situation where Alice plays her part, but Bob is malicious from the start? Here we are not in the scope of a situation 
(covered by the operator $\atlop{A,B}$) where both Alice and Bob cooperate. The formula is therefore silent on this situation, and leaves it 
completely open whether Alice can recover if this happens. In attempting to constrain the situations ensuring recovery, we have gone too far. 
What we need instead is that Alice is protected from all possible behaviours of Bob, just based on the fact that Alice has been playing her part 
to cooperate with Bob.  

To express this, we seem to need a richer language with the expressiveness to name stategies. 
This expressiveness was introduced in a class of logics known as strategy logic \cite{CHP10}. 
This class of logics contain operators similar to the following.\footnote{Since our focus at this point is just to identify what kind of expressiveness
we need, we avoid a commitment to any specific logic, or going into formal details of the semantics.}
\begin{itemize} 
\item 
$\exists \alpha\semc p (\phi)$, expressing that there exists a strategy (named $\alpha$) for player $p$, such that 
formula $\phi$ holds. 
\item $\nec \phi$, expressing that whatever strategy any player plays from the present state onwards,
$\phi$ holds, 
\item 
$p \plays \alpha$, expressing that from the present moment onwards, player $p$ plays strategy $\alpha$. 
\end{itemize} 
Note that, when $G$ is the set of agents $\{p_1, \ldots, p_n\}$, the alternating temporal logic formula 
$\atlop{G} \phi$ is equivalent to 
\[ \exists \alpha_1\semc p_1 \ldots \exists \alpha_n \semc p_n \nec ( 
(\bigwedge_{i=1 \ldots n} p_i \plays \alpha_i) \rimp \phi )~.\] 

Using these constructs, we can write the following: 
\[
\exists \alpha\semc A
\exists \beta\semc B 
 \left (I \rimp  \left (
\begin{array}[c]{l}
\nec ((A\plays \alpha \land B\plays\beta) \rimp \Eventually F) \land\\ 
\nec (A\plays\alpha \rimp (\recoverable_A \awaiting  F)) ~\land \\
\nec (B \plays \beta \rimp (\recoverable_B \awaiting  F)) 
\end{array} \right) \right)
\]
Intuitively, here $\alpha$ and $\beta$ are Alice and Bob's strategies when they cooperate. 
The first of the three inner formulas therefore says that the final condition $F$ is guaranteed to eventually hold 
if they both cooperate. The following two formulas say that, each is assured, based just on the fact that 
they are playing their cooperative strategy, and independent of what the other agent (or any third party) does,  
that until such time as the final condition $F$ holds, they have retained the ability to recover one or the other asset
by executing some strategy. 

This significantly improves our specification of the atomic swap. However, there remains a subtle issue. 
Recall that the  formula $\recoverable_A$ says that there is a strategy by which $A$ can eventually recover an 
asset. It does \emph{not} say that $A$ knows what this strategy is. This strategy may also vary from moment to moment, 
while $A$ is waiting to determine whether Bob is cooperating, and has not yet decided to attempt to recover an asset. 
There is a difference between knowing that there exists something satisfying some property, and there existing something
that is known to satisfy some property! This distinction is well known in the philosophy literature as the 
\emph{de dicto} - \emph{de re} distinction.  

Particularly once we introduce cryptographic constructs, this distinction matters. 
Suppose we modify the smart contract of Figure~\ref{fig:sc1} by adding a 256 bit integer variable 
$\mathtt{hash}$, enable Bob (but not Alice) to set this variable at most once using a new function $\mathtt{setHashB}$, and modify Alice's operation 
$\mathtt{cancel}$ so that it has an effect only if Alice provides a string that hashes under the 
SHA-256 hash function to the value of $\mathtt{hash}$. 
The modifications are shown in Figure~\ref{fig:sc2}.

\begin{figure} 
 \begin{verbatim}  
hash : Int256

initialise { depositedA := False ;  depositedB := False ;  hash := 0}  

setHashB(h:Int256)  { if sender = Bob and hash = 0 then { hash:= h }  }

cancelA(x:String) { if depositedA  and sha256(x) = hash 
                       then { depositedA := False; send(a,A) }  }  
 \end{verbatim} 
 \caption{\label{fig:sc2}A variant of actions in the atomic exchange smart contract} 
 \end{figure} 

It is now the case that while cooperating, and before $F$ holds, Alice will have always have a strategy for recovery of an asset: the 
strategy is to call $\mathtt{cancelA}$ with a string whose hash is $\mathtt{hash}$. In case Bob manages to update the value from 
$0$ to a new value (which he can do at most once), Alice should repeat this with the inverse of the new value. 
However, cryptographic hash functions like SHA-256 have the property of requiring large amounts of computation to compute
an inverse. Therefore,  this strategy may not be immediately known to Alice, and she may need to conduct an infeasible amount of computation to 
discover such a string! 
(Alternately she could call $\mathtt{cancelA}$ with all strings in succession until one work,  but this again 
requires a very large expected effort, as well as a very high expected cost in transaction fees in a blockchain 
setting that charges her for each call.)
It would therefore be misleading to say that Alice is protected against Bob's malice because she 
has a strategy for recovery of an asset. 

There exist proposals for strategy logics that add operators that talk about what agents know as well as what they 
are able to do (\cite{HM18} is one recent work, with references to others), but it is not clear that they resolve this particular problem.\footnote{For 
one thing, according to the usual semantics of knowledge, agents have unbounded computational powers, so these logics
would in fact (contrary to intuition) state that Alice knows a strategy  for recovering an asset when this requires solving a hash puzzle.} 
Instead, we resolve it by making the implicitly quantified strategy in $\recoverable_i$ explicit, and 
moving it to the the front of the formula. For $i= A,B$, write $\recoverable_i(\alpha)$  
for the formula 
\[\nec( i \plays \alpha \rimp \Eventually (\holds(i,a) \lor  \holds(i,b)) ~. \]
This says that strategy $\alpha$ can be played by agent $i$ to recover one of the two assets. 
Then we can overcome the issue of agent's not knowing the correct current value of a strategy
that changes from state to state by means of the alternate specification
\[
\exists \alpha,\alpha_R\semc A
\exists \beta,\beta_R \semc B ( \Swap(\alpha, \alpha_R,\beta,\beta_R)) \] 
where $\Swap(\alpha, \alpha_R,\beta,\beta_R)$ is the formula 
\[ 
 I \rimp  \left (
\begin{array}[c]{l}
\nec ((A\plays \alpha \land B\plays\beta) \rimp \Eventually F) \land\\ 
\nec (A\plays\alpha \rimp (\recoverable_A(\alpha_R) \awaiting  F)) ~\land \\
\nec (B \plays \beta \rimp (\recoverable_B(\beta_R)  \awaiting  F)) 
\end{array} \right) 
\]
Here $\alpha$ and $\beta$ are Alice and Bob's strategies when they cooperate, 
and $\alpha_R$ and $\beta_R$ are the strategies that Alice and Bob can play to recover an asset should 
they decide the other is not cooperating. We have finessed the de dicto - de re issue by 
identifying specific strategy pairs $(\alpha,\alpha_R)$ and $(\beta,\beta_R)$.  If the specification is valid, witnessed by
these strategies, then Alice and Bob, provided with these strategy pairs, will know  what to do. 
 
\section{Model checking an atomic swap} \label{sec:mc} 

The specification we have developed above uses quantification over strategies. 
This is appropriate for an abstract specification of a swap contract: it gives maximum flexibility to 
the implementer to decide what concrete operations the smart contract should provide, and how 
the agents should use those operations in order to achieve the desired effect. 

On the other hand, the complexity of automated verification is, in general, higher for logics containing quantification over strategies than it 
is for the less expressive temporal logics on which they build. Quantification over strategies involves a version of the \emph{synthesis} problem, 
that of finding a strategy satisfying a given property, from a potentially infinite set of possibilities. 
In multi-agent settings, this problem is computationally undecidable in its most general forms \cite{PR90}, and decidable cases  tend to 
be highly restricted, but still computationally infeasible.

However, for our purposes, this complexity can be avoided. While the specification asks for the existence  of a set of strategies, 
an implementation can be asked to deliver  not only a smart contract satisfying the specification, but also a set of concrete 
strategies $\alpha, \alpha_R,\beta,\beta_R$ that witness the claim of existence. 
Verification of the claim amounts to checking that the formula $\Swap(\alpha, \alpha_R,\beta,\beta_R)$ holds for these strategies. 
This formula has no remaining strategic operators, but only temporal operators, making it amenable to automated
verification using model checking technology. 

We demonstrate this by verifying the smart contract of Figure~\ref{fig:sc1} with respect to a set of strategies for Alice and Bob. 
The strategies are that each either cooperates, transferring their asset to the smart contract and calling \verb+finalise+ when it holds
both, or recovers by calling their \verb+cancel+ action. We have used the model checker 
MCK \cite{GM04},  but expect that it could be carried out without difficulty in other model checking systems.\footnote{The principal novelty of MCK
is that it supports specification that concern the knowledge of agents, but we do not use this facility in the present paper.} 
The complete text of our MCK model is given in Appendix~\ref{app:onchain}.\footnote{The MCK scripts discussed in this paper 
are also available as examples in the MCK web app at \url{http://cgi.cse.unsw.edu.au/~mck/mckform/}.}
MCK verifies the formulas in under a second.

A few issues arise in this exercise. One is how to represent the smart contract and the agent strategies. 
MCK provides a modelling language that separates systems into an environment (similar to the concurrent game structures used in the 
semantics of alternating temporal and strategy logics)  and protocols executed by each of a number of agents.  
We use the environment to model the smart contract, and encode the possible strategies in the agent protocols. 
The environment's encoding of the smart contract is a straightforward transcription of the code of Figure~\ref{fig:sc1}. 

Since the specification involves agents potentially switching from one strategy to another during the course of a run, 
we do not model each strategy as an individual protocol. Instead, we construct a single protocol for Alice that encodes
the rules for the cooperating strategy $\alpha$, the recovery strategy $\alpha_R$, as well as a strategy in which Alice chooses
her actions nondeterministically. This nondeterministic protocol  covers all possible behaviours of Alice, and 
is included in order to enable reasoning about the situation where Alice acts maliciously. A variable \verb+strategyA+ taking a value in 
the set of constants \verb+{Cooperate,Recover,Random}+ is used to select Alice's choice of strategy at each moment of time. 
This variable is assigned a value non-deterministically, and Alice's protocol then selects Alice's action in accordance with the 
corresponding strategy for Alice. We can then express ``Alice plays the cooperation strategy $\alpha$ at all times in the future''
by a formula \verb+Always(strategyA = Cooperate)+. Some minor modelling details that motivate a small change to the formulas are discussed
in the appendix. 

\section{Cross Chain Swaps} \label{sec:ccswap} 

We considered above a  swap performed in the context of a single blockchain. 
The blockchain community has also developed an approach to \emph{cross-chain} swaps, 
in which the assets to be swapped exist on different blockchain systems. 
The specification approach discussed above continues to apply, but implementing it 
in such a setting is a harder problem. If Alice and Bob's assets reside on different blockchains, 
then we can no longer have them deposit both assets into the control of a single 
smart contract - we need to use a number of interoperating smart contracts
residing on different chains. 

A solution that has been proposed \cite{HTLC,ACCT} uses a combination of hash and time locks. 
A time lock is a constraint stating that an action (e.g., pay an asset to some party) cannot
be performed until a specified time. A hash lock is an output value $y$  of a cryptographic hash function $h$, 
and requires that the caller of an action supply a value $x$ such that $h(x) = y$ in order to 
execute the action. 

Suppose now that Alice's asset $a$ resides on a blockchain $BC_A$, and Bob's asset $b$ resides on another $BC_B$. 
A protocol for executing the swap then proceeds as follows: 
\begin{enumerate} 

\item Alice generates a random value $s$ from a space large enough that Bob will not be able to 
guess it in a feasible amount of time. She keeps this secret until a later step of the protocol. 

\item Alice creates a smart contract on $BC_A$ with hash lock $h(s)$ and time lock $t_A$
such that if Bob provides $s$ before $t_A$ then the asset $a$ will be transferred to Bob, else $a$ will 
revert to Alice after $t_A$. 

\item Since Bob, can now see the value of $h(s)$, he is able to create a smart contract on $BC_B$ 
with hash lock $h(s)$ and time lock $t_B$ such that if Alice provides $s$ before $t_A$ then the asset $b$ 
will be transferred to Alice, else will revert to Bob after $t_B$. 

\item Before $t_B$, Alice uses $s$ to collect $b$ from Bob's smart contract. This reveals $s$ onto 
Bob's blockchain. 

\item Before $t_A$, Bob uses $s$ to collect $a$ from Alice's smart contract. 
\end{enumerate}
In order to be correct, the times $t_A$ and $t_B$ need to be selected so that the network delays in having smart contracts logged and 
actions executed on smart contracts is taken into account, and  the agents get adequate time to 
collect the assets before the timeouts. In particular, we need that $t_A$ is an adequate amount greater than the latest time $t_B$ that Alice
might collect asset $b$ and reveal the secret, to allow Bob to also collect asset $a$. 

We have also modelled this protocol in MCK, and verified that it satisfies our specification.  
(Verification runs in about 17 seconds in this case.) 
The details are given in Appendix~\ref{app:crosschain}. 
In this case, the main issue arising in the exercise is that if we are to stay in the domain of temporal logic model checking, 
it is necessary to abstract the hash lock, and the way that inhibits the performance of an action until the agent knows the 
secret. We have done so by means of a boolean variable that records whether Alice has revealed the secret in collecting
$b$. This is a simple instance of the ``perfect cryptography'' assumptions that are commonly made in model checking cryptographic 
protocols. 

A more realistic modelling would encode the hash function more  explicitly 
(e.g., as a random oracle, or more ambitiously, as actual code) 
and state the specification probabilistically. A difficulty with such a more accurate model
is that the computational cost of automated verification of probabilistic specifications is
inherently higher than that of discrete specifications (if not undecidable), 
and such an approach will tend to require semi-automated theorem proving with 
larger amounts of human input to the verification process. 
We note, however,  that there are correctness  results for the abstract ``perfect cryptography'' models of 
the kind we have used, and we suggest that this is the preferred route for automated verification of protocols
involving hash-locks.  We leave a more detailed exploration of this issue for future work. 

\section{Discussion} \label{sec:concl}

Our examination of the question of how to specify an atomic swap smart contracts suggests that while pre-condition/post-condition specifications do 
not suffice, ideas from logics with quantification over strategies are conceptually useful for stating requirements, but in practice temporal logic verification methods
may suffice for verifying concrete implementations of smart contracts, and the strategies that are used to operate them. 
We conclude by discussing some issues we have left unaddressed, which provide questions for future work.  

One issue that our specifications have not covered is the question of what is permitted to vary besides
control of the assets $a$ and $b$. It can be expected that some other things \emph{will} vary in any implementation. 
For example, in order to execute the smart contracts involved, Alice and/or Bob are likely to have to pay transaction fees, 
so their holding of crypto-currency is also likely to have to vary. It may be desirable to place some bound on their total costs.  
On the other hand, beyond this, we would not want an implementation that requires Alice and Bob to incur losses in their holdings of 
other assets. It would therefore be desirable to add a statement to the specification indicating  exactly what is allowed to 
vary in a solution. In the refinement calculus, this is often done by means of a \emph{frame specification}. For example, 
a pre-condition/post-condition specification $[\phi,\psi]_X$ asks for a program satisfying $[\phi,\psi]$, while 
modifying \emph{no more than} the variables in the set $X$. It would be desirable to add such a constraint to our specifications, 
but temporal logic and its extensions usually lack the expressive power for such constraints.
Some steps in this direction have been taken in the exploration of connection between temporal logics and separation logic
\cite{Brochenin13},  and it would be interesting to explore this direction for our context. 

In our model checking exercise, we have verified a concrete implementation that talks only about 
Alice and Bob's holdings of $a$ and $b$, so it seems reasonable that it follows that 
a frame condition stating that nothing else varies 
will be true of this implementation. However, it would be desirable to have a precise statement of this 
claim. For one thing, an implementation containing actions whose execution requires
all the available time and/or space in a block will prevent other actions being performed, so 
it is not true that a smart contract runs entirely free from interference with other contracts on a disjoint set of assets, particularly when these 
have deadlines of their own.  

Various improvements are possible to our modelling, that we have not pursued since they are orthogonal to our main objective of 
eliciting the type of specification and verification required for the smart contracts we consider. 
We have not attempted  to build an accurate model of the way that agents submit their transaction to the network, or the way that miners 
select those transactions for execution. Our models assume that agents offer an action for execution at each step, and can retract those 
actions instantaneously. In a real network there would be delays, uncertainties and transaction costs in doing so. 
Duplicate or conflicting transactions offering different transaction fees is something that may need to be incorporated in the miner model. 
In this event, it may then become relevant to consider game theoretic analysis: we expect that miners will opt to take a higher transaction fee, 
although a miner who has not yet received a message offering a higher gas price may still use a lower price transaction. 

We have considered only bilateral swaps in this paper.  Herlihy \cite{Herlihy18} generalises from two-party swap protocols  to an arbitrary number of parties 
with a set of transfers represented using a directed acyclic graph. 
His protocol uses multi-signature hash-locks, where a 
chain of signatures on the secret is required to open a hash-lock. 
To verify this protocol using temporal logic techniques would require an 
additional abstraction to be made to represent which agents have signed a particular message. 
While we do not think that this would be difficult to model, it would be an interesting direction to explore. 

Finally, what we have verified in our work is an abstract model of smart contract code rather than code in a 
native blockchain smart contract language. It would be beneficial to bridge this gap. Two approaches are possible. One is to verify 
an abstract modelling notation that comes equipped with a correctness preserving translation to 
native blockchain code. The other is to have a tool that verifies temporal specifications directly on blockchain code. 
We note however, that what needs to be verified is not just the smart contract, but also the strategies for operating it, 
and the latter are off-chain, and generally represented in some other language. A further complication is that in the 
multi-chain setting, we need to deal with multiple, possibly quite distinct smart contract languages for each chain. 
This suggests that the former route, of conducting verification using notations that abstract from chain-specific details, 
may be the cleaner route to pursue. A further benefit of this would be that it could provide automated support for correct 
implementation of details that are abstracted by the framework for verification purposes, such as the hash locks and signatures.

\bibliographystyle{alpha} 
\bibliography{refs}

\appendix

\section{MCK model of an on-chain swap contract} \label{app:onchain}

The following gives MCK code for the on-chain swap contract and the strategies that 
operate it. The codes starts by declaring a number of types and defining the environment, 
by listing a number of variables that make up the environment state. 
The MCK statement \verb+init_cond+ is used to constrain initial states of the environment: 
all states that satisfy the condition in this statement are initial. Note that since the variables 
\verb+strategyA+, \verb+strategyB+ are not mentioned, they may take any value in an initial state. 
The \verb+define+ statements can be understood as macro definitions. An \verb+agent+ statement 
introduces an agent and identifies the protocol that the agent uses to choose its actions. The protocol arguments in this 
statement are used to alias the protocol's parameter variables to variables in the environment. 

The \verb+transitions+ statement gives (nondeterministic) code that is used to make a single transition of the environment state, 
depending on actions that have been selected by the agents as inputs to the transition. 
In protocols, an action \verb+a+ is denoted by \verb+<<a>>+. 
In the environment, there is a corresponding atomic proposition (i.e., boolean variable)  $i.a$ for each agent $i$, 
representing the fact that the agent is performing action $a$ in the current state transition. For example, in our model, 
Alice has an action \verb+Deposit+, and there is a corresponding atomic proposition \verb+Alice.Deposit+. 

A statement \verb+if ... fi+ is a Dijkstra style nondeterministic guarded choice statement. 
A  statement  $[[vars|\phi]]$ also nondeterministic. 
The list of variables $vars$ in its left hand side are all the variables whose value may change when running this statement. 
The formula $\phi$ in the right hand side gives a condition on a pair $(s,s')$, where $s,s'$ are states of the environment, 
through use of state variables to refer to component $s$ and ``primed'' version of state variables to refer to $s'$. 
Started in a state $s$, this statement can make a transition to any state $s'$ (in which only the variables in \verb+vars+
may have been changed) such that the pair $(s,s')$ satisfies the formula $\phi$. For the particular application of this statement, 
\verb+[[ strategyA,strategyB, turn | True ]]+ has the effect of non-deterministically selecting values for each of the listed variables.

The model schedules one of the agents to have its choice of action executed in each transition. 
The variable \verb+turn+ indicates the scheduled agent. 
Each statement \verb+fairness+ gives a condition that must be true infinitely often in each run. Thus 
\verb+fairness = turn == AliceP+ says that Alice is treated fairly in the sense that she gets infinitely many turns in each run (but
she may have to wait an arbitrary amount of time between turns).

This is followed by statements \verb+spec_obs+, which give the formulas to be model checked. (The component \verb+obs+ in this 
keyword indicates the semantics to be used for knowledge modalities - since there are none in our specification, this is irrelevant here.) 
The formulas are those from the body of the paper, using MCK's notation  (consistent with the literature)  for the temporal operators: 
$A$ says ``on all runs from the present state", $G$ says ``at all future times", $F$ says ``at some future time", and 
$\alpha U \beta$ says $\alpha$ holds up until some future time where $\beta$ holds.

The specifications contain one minor deviation from the formulas given above: 
we need a reformulation of the formula $$A\plays\alpha \rimp (\recoverable_A(\alpha_R) \awaiting  F)~.$$ 
(A revised version of the MCK scripting under development 
will help to avoid this deviation.) The reason for this is that we encode agents' choice of strategy as a state variable, and the specifics of 
MCK's execution model, in which, at each transition, 
the state is used to select agent actions which are then used as inputs to the environment transition code. 
On a run  where $A\plays \alpha$, at a point where $A$ switches to playing $\alpha_R$, she is already 
committed to playing $\alpha$ for the next step. Thus, rather than saying that $A$ always plays $\alpha$, 
we talk about what happens when $A$ switches to playing $\alpha_R$ in a situation where she has always played 
$\alpha$ in the past. For this, it is  convenient to also build in a variable \verb+playedCoopA+ that tracks whether Alice cooperated 
at all times in the past.

{\small 
\begin{verbatim} 

type Strategy = {Cooperate,Recover,Random} 
type Holder = {AliceH,BobH,Contract,Other} 
type Player = {AliceP,BobP}

done : Bool 

depositedA : Bool 
holdera : Holder 

depositedB : Bool 
holderb: Holder

strategyA : Strategy
strategyB : Strategy

turn : Player

{- the following track whether a player has always 
   cooperated up to the present time -} 
playedCoopA : Bool
playedCoopB : Bool 


define swapped = holdera == BobH /\ holderb == AliceH

define alice_safe = holdera == AliceH \/ holderb == AliceH

define bob_safe = holdera == BobH \/ holderb == BobH

init_cond = holdera == AliceH /\ holderb == BobH /\
            playedCoopA /\ playedCoopB /\ 
            neg (done \/ depositedA \/ depositedB ) 

agent Alice "player" (strategyA,depositedA,depositedA,depositedB)  

agent Bob "player" (strategyB,depositedB,depositedA,depositedB) 


transitions
begin 
 -- process Alice or Bob's action, if it is their turn
 if 
  turn == AliceP -> 
    if    
	   Alice.Deposit /\ holdera == AliceH -> 
		  begin depositedA := True ; holdera := Contract end 

	[] Alice.Cancel /\ holdera == Contract -> 
	       begin depositedA := False; holdera := AliceH end

	[] Alice.Finalize /\ depositedA /\ depositedB -> 
		  begin depositedA := False ; depositedB := False ; 
		            holdera := BobH ; holderb := AliceH end

        -- GiveToOther means give away one of your assets
        [] Alice.GiveToOther /\ holdera == AliceH -> holdera := Other
        [] Alice.GiveToOther /\ holderb == AliceH -> holderb := Other 

	[] otherwise -> skip 
    fi 
[] turn == BobP -> 
   if 
       Bob.Deposit /\ holderb == BobH ->
          begin depositedB := True ; holderb:= Contract end 

    [] Bob.Cancel /\ holderb == Contract /\ depositedB -> 
           begin depositedB := False; holderb := BobH end

    [] Bob.Finalize /\ depositedA /\ depositedB -> 
           begin depositedA := False ; depositedB := False ;
                    holdera := BobH ; holderb := AliceH end 

    -- GiveToOther means give away one of your assets
    [] Bob.GiveToOther /\ holdera == BobH -> holdera := Other
    [] Bob.GiveToOther /\ holderb == BobH -> holderb := Other 


    [] otherwise -> skip 
   fi 
 fi ;

-- update the record of whether a player has always cooperated
if strategyA /= Cooperate -> playedCoopA := False fi ;
if strategyB /= Cooperate -> playedCoopB := False fi ;

 -- nondeterministically select each player's strategy, and whose turn it is
 [[ strategyA,strategyB, turn | True ]]  
end 


-- Both Alice and Bob get infinitely many turns in each run 
fairness = turn == AliceP
fairness = turn == BobP



spec_obs = "If Alice and Bob always play Cooperate, 
                     then eventually the swap is successful" 
   A( (G(strategyA == Cooperate /\ strategyB == Cooperate)) => F swapped) 



{- In the following, note "A awaiting B" is "(A until B) or (always A and not B)" -}

spec_obs = "If Alice has always cooperated to now, then she can get an asset back 
                     by playing recover, awaiting such time as the swap has occurred"
  A( ( (playedCoopA => ((G strategyA == Recover) => F alice_safe)) U swapped )
     \/  
     G ( (playedCoopA => ((G strategyA == Recover) => F alice_safe))
         /\ 
         neg swapped ))


spec_obs = "If Bob has always cooperated to now, then he can get an asset back 
                     by playing recover, awaiting such time as the swap has occurred"
  A( ( (playedCoopB => ((G strategyB == Recover) => F bob_safe)) U swapped )
     \/  
     G ( (playedCoopB => ((G strategyB == Recover) => F bob_safe))
         /\ 
         neg swapped ))



{- Note that because a finalise operation might race with a cancel operation, recovery 
   of your own asset is not guaranteed! -} 

spec_obs =  "(FALSE) Alice is always able to ensure that she will eventually hold asset a, 
                      by playing strategy Recover"
  A( ( (playedCoopA => ((G strategyA == Recover) => F holdera == AliceH )) U swapped )
     \/  
     G ( (playedCoopA => ((G strategyA == Recover) => F holdera == AliceH))
         /\ 
         neg swapped ))



spec_obs =  "(FALSE) Bob is always able to ensure that he will eventually hold asset b, 
                      by playing strategy Recover"
  A( ( (playedCoopB => ((G strategyB == Recover) => F holderb == BobH )) U swapped )
     \/  
     G ( (playedCoopB => ((G strategyB == Recover) => F holderb == BobH))
         /\ 
         neg swapped ))




protocol "player" (strategy : Strategy, deposited : Bool, 
                            depositedA : Bool ,depositedB : Bool) 
 
begin
do 
    strategy == Cooperate /\ neg deposited  -> <<Deposit>> 
[]  strategy == Cooperate /\ depositedA /\ depositedB  -> <<Finalize>> 

[]  strategy == Recover /\ deposited  -> <<Cancel>>

[]  strategy == Random -> <<Deposit>>
[]  strategy == Random -> <<Cancel>>
[]  strategy == Random -> <<Finalize>>
[]  strategy == Random -> <<Skip>>
[]  strategy == Random -> <<GiveToOther>>

[]  otherwise -> <<Skip>> 
od 
end 

\end{verbatim} 

}

\section{MCK model of  a cross-chain atomic swap protocol} \label{app:crosschain}

The following is MCK code for a cross-chain atomic swap protocol using hashed time locks (with the hash checking and knowledge of the 
secret abstracted using a single boolean variable rather than a large range of possible values for the secret). 
Because of the timeouts, we are able to combine the Cooperate strategy and the Recover strategy into 
a single strategy for each player, in which they perform the recovery action in case their smart contract 
times out without the other player having claimed the asset. This also allows for a 
simpler, stronger specification statement concerning recovery. All the specifications in the following script are verified to hold.  

{\small 
\begin{verbatim}


type Strategy = {Cooperate,Recover,Random} 
type Holder = {AliceH,BobH,ContractA,ContractB,Other} 
type Player = {AliceP,BobP}

type Time = {0..20} 

{-
   The secret x is abstracted - type Secret is used to model 
   whether a player knows the secret.  If so, their Claim action is effective
-}    
type Secret = {Known,Unknown} 


holdera : Holder 
holderb: Holder

strategyA : Strategy
strategyB : Strategy

turn : Player
time : Time 

{- the following model each player's view of Alice's secret -} 
viewSecretA : Secret
viewSecretB : Secret


{- Whether an asset has been paid into the contract.
   These are used to prevent paying the asset into the contract
   once again in the event it is recovered. -} 
depositedA : Bool
depositedB : Bool 

timeoutA : Time
timeoutB : Time 


-- the following track whether a player has always cooperated up to the present time 
playedCoopA : Bool
playedCoopB : Bool 


define swapped = holdera == BobH /\ holderb == AliceH

define alice_safe = holdera == AliceH \/ holderb == AliceH

define bob_safe = holdera == BobH \/ holderb == BobH


init_cond = holdera == AliceH /\ holderb == BobH /\
            viewSecretA == Unknown /\ viewSecretB == Unknown /\ 
            playedCoopA /\ playedCoopB /\ time == 0 /\
	    -- (constant) timeouts for Alice and Bob's contracts  
            timeoutA == 8 /\ timeoutB == 6  /\ turn == AliceP /\
	    neg (depositedA \/ depositedB) 





agent Alice "playerA" (strategyA,holdera,holderb,viewSecretA,
                                   time,timeoutA,depositedA)  

agent Bob "playerB" (strategyB,holderb,holdera,viewSecretB,
                                  time,timeoutB,depositedB) 


transitions
begin 
 -- process Alice or Bob's action, if it is their turn
 if 
    turn == AliceP -> 
    if
       Alice.Generate -> viewSecretA := Known 
    [] Alice.Deposit /\ holdera == AliceH ->
              begin holdera := ContractA ; depositedA := True end  

    -- when Alice claims from Bob's contract, she reveals the secret to Bob   
    [] Alice.Claim /\ viewSecretA == Known  -> 
         begin
             if holderb == ContractB -> holderb := AliceH fi ; 
             viewSecretB := Known 
         end

    [] Alice.Recover /\ holdera == ContractA /\ time >= timeoutA ->
         holdera := AliceH 

        -- GiveToOther means give away one of your assets
    [] Alice.GiveToOther /\ holdera == AliceH -> holdera := Other
    [] Alice.GiveToOther /\ holderb == AliceH -> holderb := Other 

    [] otherwise -> skip 
    fi 
[] turn == BobP -> 
   if 
      Bob.Deposit /\ holderb == BobH ->
         begin holderb:= ContractB ; depositedB := True end

     [] Bob.Claim /\ viewSecretB == Known /\ holdera == ContractA ->
            holdera := BobH 

     [] Bob.Recover /\ holderb == ContractB /\ time >= timeoutB ->
             holderb := BobH

       -- GiveToOther means give away one of your assets
     [] Bob.GiveToOther /\ holdera == BobH -> holdera := Other
     [] Bob.GiveToOther /\ holderb == BobH -> holderb := Other 

     [] otherwise -> skip 
   fi 
 fi ;

-- update the record of whether a player has always cooperated
if strategyA /= Cooperate -> playedCoopA := False fi ;
if strategyB /= Cooperate -> playedCoopB := False fi ;

-- Assume the two chains are synchronised, modelled by round robin scheduling
if     turn == AliceP -> turn := BobP
   [] turn == BobP -> turn := AliceP
fi ;

-- randomly select the strategy followed by each player at each step.  
[[ strategyA,strategyB | True ]]  ;

-- tick the clock 
time := time + 1 
end 


spec_obs = "If Alice and Bob always play Cooperate, 
                     then eventually the swap is successful" 
   A( (G(strategyA == Cooperate /\ strategyB == Cooperate)) => F swapped) 

spec_obs = "If Alice always cooperates, she is always eventually safe" 
  A( (G strategyA == Cooperate) => G F alice_safe )

spec_obs = "If Bob always cooperates, he is always eventually safe" 
  A( (G strategyB == Cooperate) => G F bob_safe )



protocol "playerA" (strategy : Strategy,
		    holder_mine : Holder,
		    holder_other : Holder,
		    viewSecret : Secret,
		    time : Time,
		    timeout: Time,
		    deposited: Bool)
 
begin
do 

    strategy == Cooperate -> 
      if  -- generate a secret (which becomes Known) 
          viewSecret == Unknown -> <<Generate>>
      -- unless you have done so before, deposit your asset 
      []  viewSecret == Known /\ holder_mine == AliceH /\ 
              neg deposited -> <<Deposit>>
      -- make a claim as soon as you know the secret and Bob has deposited
      []  viewSecret == Known /\ holder_other == ContractB -> <<Claim>>
      -- try to recover your own asset at the earliest time after the timeout
      []  holder_mine == ContractA /\ time >= timeout -> <<Recover>>
      fi 

[]  strategy == Random -> <<Generate>>
[]  strategy == Random -> <<Deposit>>
[]  strategy == Random -> <<Claim>>
[]  strategy == Random -> <<Recover>>
[]  strategy == Random -> <<Skip>>
[]  strategy == Random -> <<GiveToOther>>

[]  otherwise -> <<Skip>> 
od 
end




protocol "playerB" (strategy : Strategy,
		    holder_mine : Holder,
		    holder_other : Holder,
		    viewSecret : Secret,
		    time : Time, 
		    timeout: Time, 
                    deposited : Bool ) 
 
begin
do

    strategy == Cooperate  ->
      if  -- once you see Alice has made her deposit, make yours 
          holder_mine == BobH /\ holder_other == ContractA /\ 
          neg deposited -> <<Deposit>>

          -- make a claim as soon as you know the secret
      []  viewSecret == Known /\ holder_other == ContractA -> <<Claim>> 

         -- try to recover your own at the earliest possible time 
      []  holder_mine == ContractB /\ time >= timeout -> <<Recover>>
      fi 

[]  strategy == Random -> <<Deposit>>
[]  strategy == Random -> <<Claim>>
[]  strategy == Random -> <<Recover>>
[]  strategy == Random -> <<Skip>>
[]  strategy == Random -> <<GiveToOther>>

[]  otherwise -> <<Skip>> 
od 
end 

\end{verbatim}
}

\end{document}